\documentclass[a4paper,twocolumn,10pt]{quantumarticle}
\pdfoutput=1

\usepackage[utf8]{inputenc}
\usepackage[T1]{fontenc}

\usepackage{graphicx}
\usepackage{physics}
\usepackage{amsmath}
\usepackage{amsfonts}
\usepackage{amssymb}
\usepackage{multirow}
\usepackage{xcolor}
\usepackage{comment}
\usepackage[ruled,vlined,linesnumbered]{algorithm2e}
\usepackage{hyperref}
\usepackage{cleveref}

\usepackage{lipsum}

\SetKwFor{While}{while}{}{end while}
\SetKwFor{ForEach}{foreach}{}{end foreach}

\begin{document}

\title{On the complexity of quantum link prediction in complex networks}

\author{João P. Moutinho}
\affiliation{Instituto Superior Técnico, Universidade de Lisboa, Portugal}
\affiliation{Instituto de Telecomunicações, Lisboa, Portugal}
\altaffiliation{Corresponding author: joao.p.moutinho@tecnico.ulisboa.pt}
\author{Duarte Magano}
\affiliation{Instituto Superior Técnico, Universidade de Lisboa, Portugal}
\affiliation{Instituto de Telecomunicações, Lisboa, Portugal}
\author{Bruno Coutinho}
\affiliation{Instituto de Telecomunicações, Lisboa, Portugal}


\begin{abstract}
\noindent Link prediction methods use patterns in known network data to infer which connections may be missing. Previous work has shown that continuous-time quantum walks can be used to represent path-based link prediction, which we further study here to develop a more optimized quantum algorithm. Using a sampling framework for link prediction, we analyze the query access to the input network required to produce a certain number of prediction samples. Considering both well-known classical path-based algorithms using powers of the adjacency matrix as well as our proposed quantum algorithm for path-based link prediction, we argue that there is a polynomial quantum advantage on the dependence on $N$, the number of nodes in the network. We further argue that the complexity of our algorithm, although sub-linear in $N$, is limited by the complexity of performing a quantum simulation of the network's adjacency matrix, which may prove to be an important problem in the development of quantum algorithms for network science in general.
\end{abstract}

\maketitle

\section{Introduction}

Complex networks provide a common framework to study different complex systems \cite{barabasi2016network}. Representing agents as nodes and interactions as links is a general enough description to fit many real systems, such as protein-protein interaction networks, social networks, transportation networks, electrical grids, and many others. Over the years, network science has lead to the realisation that these different systems share many common structural properties, and it is often possible to gain system-specific insights through general network-based problems and tools \cite{barabasi1999emergence, albert2002statistical, dorogovtsev2002evolution, newman2003structure, boccaletti2006complex}. In the study of human disease, for example, the sub-field of Network Medicine has emerged from the success of network-based tools in problems such as the prediction of drug-combinations and cancer-driver genes \cite{horn2018netsig, cheng2019network, barabasi2011network}.

One network problem with multidisciplinary applications is that of link prediction, which aims to infer new or unobserved links from a network based on its current or known topology \cite{Liben:2007, Lu:2011, wang2015link, zhou2021progresses}. In biological networks, link prediction has important applications in the identification of unknown protein-protein interactions \cite{kovacs2019network}, or aiding in the mapping of large scale neural networks \cite{yang2015link, cannistraci2013link}, which better our understanding of human biology. In online social and commerce networks it can be used to suggest new friendships between users \cite{Liben:2007, wang2015link} or make product recommendations, increasing customer retention \cite{talasu2017link, huang2005link, daminelli2015common}.

Link prediction is often a computationally intensive task. To make informed predictions, methods evaluate a certain score function over the whole set of potentially missing links to identify the few that stand out as the best predictions \cite{zhou2021progresses}. Recently, it has been suggested that the usage of quantum computers may help speed-up link prediction by sampling new links from a quantum walk evolution encoding the score values \cite{moutinho2021quantum}. This result was one of the first examples of a quantum algorithm developed based on network science insights, with applications ranging from social network analysis to network medicine problems. The importance of quantum computing applied to network medicine and link prediction was further discussed in \cite{maniscalco2022quantum} and \cite{goldsmith2022link}, respectively. Previous works in quantum walk algorithms have also tackled important network problems such as graph transversal \cite{farhi1998quantum, childs2003exponential} and marked node search \cite{childs2004spatial, chakraborty2016spatial, apers2022quadratic}.

In this work we further study the problem of link prediction and the path-based approach using Continuous-Time Quantum Walks (CTQW). First, we describe a sampling-based framework for link prediction, allowing us to make more precise resource comparisons between classical and quantum sampling-based algorithms. Previous results in classical algorithms have suggested that sampling-based algorithms may be advantageous for network-problems, including link prediction \cite{seshadhri2013triadic, jha2015path, ballard2015diamond}, which we can directly compare to. For this purpose, we use the standard benchmark of query complexity under a common input model.

Second, we provide an improved version of the quantum algorithm for link prediction initially put forward in Ref.\ \cite{moutinho2021quantum} which allows links to be sampled globally from the network. Our proposed quantum algorithm produces link prediction samples with comparable precision to the studied classical algorithms by querying the input $\tilde{O}(k_\text{max})$ times per sample where $k_\text{max}$ is the largest node degree in the network. In contrast, the classical algorithms studied typically require the full input to be queried for any number of samples, which naturally scales as $\tilde{O}(N)$, the number of nodes in the network. Using a scale-free model for complex networks \cite{barabasi2016network} the growth of the largest degree in the network can be described as $k_\text{max}\sim O\left(N^\frac{1}{1-\gamma}\right)$, with $\gamma>2$. Thus, we argue that our quantum algorithm for link prediction achieves a polynomial speedup over the classical case in the input query complexity for a fixed number of samples.

Furthermore, we argue that the quantum complexity of our link prediction algorithm is limited by the complexity of performing a quantum simulation of the network's adjacency matrix. Finding an efficient quantum simulation algorithm for complex networks will have strong implications in the development of efficient quantum algorithms for network science problems in general. A first step towards this goal was recently demonstrated in Ref.\ \cite{magano2022simulation}, where an efficient quantum simulation algorithm was developed for sparse networks with a few densely connected nodes.

We organize our work as follows: in Section \ref{sec:preliminaries} we provide all definitions and necessary background for our work. In Section \ref{sec:classicallp} we discuss relevant classical sampling-based link prediction algorithms and study their complexity. In Section \ref{sec:quantum} we provide our improved quantum algorithm for link prediction and study its complexity. In Section \ref{sec:discussion} we further discuss our results, conclusions, and future work.

\section{Preliminaries}
\label{sec:preliminaries}
\subsection{Notation}
We consider data organized in a simple, undirected and unweighted graph $\mathcal{G}(V,\,E)$, where $V$ is the set of nodes with size $N=|V|$ and $E$ is the set of links, with size $|E|$. For each node $v\in V$ we denote $\Gamma(v)$ as the set of nodes neighbouring $v$, and $\Gamma_l(v)$ as the $l$-th neighbour of $v$. The degree of $v$ is defined as $k_v = |\Gamma(v)|$, and the average degree over the network is defined as $k_\text{av}=2|E|/N$. The adjacency matrix $A\in\mathbb{R}^{N\times N}$ is such that $A_{ij}=A_{ji}=1$ if $(i,j)\in E$, and 0 otherwise.

We use the standard "big $O$" notation for asymptotic upper bounds. Given two functions $f$ and $g$ from $\mathbb{R}$ to $\mathbb{R}$ we say that $f=O(g)$ if there exists a constant $C$ such that for any $x$ greater than a threshold $x_0$ we have $f(x)<Cg(x)$. We further use the $\tilde{O}$ notation when omitting poly-logarithmic dependencies.

\subsection{Link Prediction}
Link prediction is the general problem of inferring new or unobserved links from data organized in a networked structure. That is, given an adjacency matrix $A$ of a complex network, we wish to select which unconnected pairs of nodes $(i, j)$, i.e., pairs where $A_{ij}=0$ and $i\neq j$, are the most likely to form a new link. The underlying assumption here is that there is enough information about the organizing principles of $\mathcal{G}$, or that this information can be correctly identified in its structural patterns, to make this inference with good precision. By quantifying this structural information, link prediction methods assign a prediction score $p_{ij}$ to every unconnected pair such that the probability of a prediction being correct is proportional to $p_{ij}$. However, most of the information about the score distribution is ultimately discarded, as the goal is to identify which links have the highest score. As such, it may be useful to design algorithms that can efficiently sample links according to their score distribution without outputting the full distribution.

We note also that we are considering no extra structure on the input besides the information contained in the adjacency matrix, i.e., which links are connected, $A_{ij}=1$, or which are not connected, $A_{ij}=0$. Other definitions of link prediction may consider extra structure on the input by separating links between measured or unmeasured. A measured link is a link which is known to be connected or disconnected, and an unmeasured link is unknown. In that scenario, the objective of link prediction is to evaluate the set of unmeasured links to infer which should be 1 or 0. In this work we consider the first scenario, where all unconnected links are assumed to be potentially missing, as that is the information available in most datasets. Nevertheless, our work can be directly extended to the second scenario by restricting the class of links which constitute a useful prediction from the set of unconnected links to the set of unmeasured links.

Link prediction is a widely studied problem in network science, and several approaches exist including machine learning techniques \cite{al2006link, ghasemian2020stacking}, stochastic block models \cite{guimera2009missing} or global perturbation methods \cite{lu2015toward}. A recent and comprehensive review on the topic can be found in Ref.\ \cite{zhou2021progresses}. In this work we focus on a popular class of link prediction methods where the predictions are made based on specific path structures between nodes \cite{zhou2021progresses, zhou2021experimental, Liben:2007, Zhou:2009, kovacs2019network, cannistraci2013link, Cannistraci:2018, muscoloni2022adaptive, pech2019link}, which have been shown to be competitive with other approaches in prediction precision \cite{zhou2021progresses, zhou2021experimental, muscoloni2022adaptive, muscoloni2021short}.

\subsection{Path-Based Link Prediction}
\label{sec:pathlp}

The overall mathematical structure of different path-based link prediction methods varies widely, but a common feature is the quantification of each prediction score $p_{ij}$ through the number of paths of a specific length between $i$ and $j$. For a given adjacency matrix $A$, each entry $(i, j)$ of $A^k$ represents the number of paths of length $k$ between $i$ and $j$. Initial results in path-based link prediction suggested that methods based on paths of length two are a simple way to quantify similarity between nodes \cite{Liben:2007, Zhou:2009}, 
\begin{equation}
    P\sim A^2,
\end{equation}
where $P$ is the matrix of prediction scores $p_{ij}$. These methods are often associated with social networks, but have been applied to all types of networks with varying results. More recently it was shown that link prediction methods scoring links based on direct similarity do not perform well in protein-protein interaction networks due to the fact that proteins often connect based on neighbour similarity principles \cite{kovacs2019network}, i.e., proteins connect to proteins that have neighbours that are similar to themselves. The authors in \cite{kovacs2019network} suggested that any link prediction method based on direct similarity can be extended to a neighbour similarity method by taking $P_\text{neighbour}=A.P_\text{direct}$. For the simple case of scoring links based on length two paths this implies that
\begin{equation}
    P\sim A^3.
\end{equation}
With this simple extension the authors proposed a link prediction method that was substantially better in the prediction of protein-protein interactions.

Other works have described direct similarity and neighbour similarity methods based on paths of even length and odd length. In \cite{pech2019link} a linear optimization method was proposed to predict links based on a linear combination of odd powers of $A$. The quantum algorithm initially proposed in \cite{moutinho2021quantum} uses the real and imaginary part of $e^{-iAt}$ to represent even and odd based predictions. In these works, and also in an extensive review of path-based methods \cite{zhou2021experimental}, path-based link prediction has been tested in a wide range of complex networks and several examples have been identified where either direct similarity or neighbour similarity methods tend to perform better.

Our focus in this work will be to further explore the quantum walk representation of direct and neighbour similarity for link prediction, as initially proposed in Ref.\ \cite{moutinho2021quantum}, providing a more efficient algorithm to implement it, and compare it to the simplest classical representations given by $A^2$ and $A^3$ in terms of query complexity. To do so, we will use a link prediction sampling framework, described next.

\subsection{Sampling Path-Based Predictions}

As discussed, the link prediction problem assumes that there is some information contained in the topological structure of the graph that can be used to infer which links are missing from the network, e.g. the number of paths of different length between nodes, computed through powers of $A$. Let us consider a general prediction method with an associated prediction matrix $P$ obtained as a function of the adjacency matrix, $P=f(A)$, which quantifies structural information that is useful for link prediction. Note that $P$ does not represent the ground-truth of the missing links, it represents only the scores $p_{ij}$ the method uses to infer which links are more likely to appear. Typically, $P$ is computed explicitly, and the scores are used to rank the predictions from best to worst. Instead, we wish to study algorithms that take as input the adjacency matrix $A$ of a given network and output samples of links $(i, j)$ following the distribution of scores $p_{ij}$ for a given $f(A)$. We write the probability of sampling a link $(i, j)$ following $f(A)$ as
\begin{equation}
\mathcal{P}[(i,j)| f]=\frac{|f(A)_{ij}|^q}{\|f(A)\|_q^q},\label{eq:cprob}
\end{equation}
with $\|.\|_q$ being the $L_{p,q}$ matrix norm for $p=q$, ensuring $\sum_{ij}\mathcal{P}[(i,j)]=1$. As we will see, the normalization is method dependent. The classical algorithms we consider sample from $f(A)=A^2$ and $f(A)=A^3$ normalized by the $L_{1,1}$ norm \cite{seshadhri2013triadic, jha2015path}, and the quantum algorithm we present samples from a distribution normalized by the $L_{2,2}$ norm.

One important thing to note is that the typical forms of $f(A)$ used not only contain information about missing links, but also about existing ones. For example, when counting the number of paths of length 3 between all pairs of nodes, the entries $(A^3)_{ij}$ encode this information irrespectively of $A_{ij}$ being 0 or 1. However, for the purpose of link prediction, the only useful predictions are those for which $A_{ij}=0$ and $i\neq j$, i.e., predictions corresponding to new links between distinct nodes. This is an important detail of the link prediction problem, as it will condition the results obtained from any algorithm sampling from $f(A)$ directly.

For any link prediction method, we may now define the probability of sampling a \textit{bad link}, i.e., a link that is useless for link prediction,
\begin{equation}
p_{\text{B}|f} \equiv \mathcal{P}[((i,j)|f)\land(A_{ij}=1\lor i=j)].
\end{equation}
The list of indices matching the condition $(A_{ij}=1\lor i=j)$ can be represented by the entries of the matrix $A+I$, which are either 0 or 1. Thus, $p_{\text{B}|f}$ can be computed by summing the probability of sampling each of these entries,
\begin{equation}
p_{\text{B}|f}=\sum_{i,j}^N(A+I)_{ij}\frac{|f(A)_{ij}|^q}{\|f(A)\|_q^q}. \label{eq:pbad0}
\end{equation}
Similarly, we can use the all-ones matrix $J$ to represent the entries that are \textit{useful} or \textit{good} for link prediction through the matrix
\begin{equation}
G=J-(A+I). \label{eq:matrixgood}
\end{equation}
Thus, the probability of a good sample is
\begin{equation}
p_{\text{G}|f}=\sum_{i,j}^NG_{ij}\frac{|f(A)_{ij}|^q}{\|f(A)\|_q^q}. \label{eq:pgood}
\end{equation}

For any algorithm sampling entries based on a link-prediction method with an underlying $f(A)$, we can expect that $O\left(1/p_{\text{G}|f}\right)$ samples will be required before an actual \textit{useful} link prediction is observed. For the remainder of the text, we omit the $f$ subscript in $p_\text{G}$, as it will be clear from context which method is being referenced. The number of samples needed to observe a \textit{correct} link prediction is harder to characterize, as that is not only dependent on the structure of $f(A)$, but also on how well $f(A)$ represents the ground-truth behind the missing links in the network, which will influence the precision of the method.

\subsection{Input Model}
\label{sec:input}

We base our comparison between classical and quantum algorithms for link prediction on both having query access to a common input model, which we now describe. We consider the \textit{general graph model} (GGM) \cite{kaufman2004tight, goldreich2017introduction}, which allows query access to the graph $\mathcal{G}=(V,\, E)$ through the following operations:
\begin{description}
\item[(1) degree query] given $v\in V$, returns the degree $k_v$;
\item[(2) neighbour query] given $v\in V$ and an integer $l$, returns the $l$-th neighbour of $v$ if $l\leq k_v$, and $\ast$ otherwise;
\item[(3) vertex-pair query] given $u,\,v\in V$, returns the adjacency matrix entry $A_{uv}$; 
\end{description}
This general model is a combination of the bounded-degree model \cite{goldreich1997property}, defined by the first and second operations, and the dense graph model, defined by the third operation \cite{goldreich1998property}. This type of input access model was initially used for complexity studies in graph property testing algorithms, and later adapted to quantum computation literature as it provides a framework where classical and quantum resources can be directly compared through the total number of queries to the input. In quantum computation literature these models are often referred to as the adjacency list model and adjacency matrix model, respectively \cite{ben2020symmetries}. A quantum extension of the general graph model can be described by defining three unitary operators $\mathcal{O}_\text{deg}$, $\mathcal{O}_\text{nei}$ and $\mathcal{O}_\text{pair}$ \cite{hamoudi2018quantum} such that
\begin{align}
    \mathcal{O}_\text{deg}\ket{v, 0}&=\ket{v, k_v},\\
    \mathcal{O}_\text{nei}\ket{v, l, 0}&=\ket{v, l, \Gamma_l(v)},\\
    \mathcal{O}_\text{pair}\ket{u, v, 0}&=\ket{u, v, A_{uv}}.
\end{align}
Our work depends on having coherent access to these oracles, such that information can be queried in a superposition, which is a standard assumption in the theoretical development of quantum algorithms. The development of QRAMs to allow such access is an active research field, and some hardware proposals have been put forward \cite{giovannetti2008quantum}. Nevertheless, their practical realization still faces significant challenges.

Finally, we consider that each of the described queries, either classical or quantum, counts as $O(1)$ in the query complexity. A classical query learns a piece of information about the input, which can be stored in a classical register and does not need to be repeated. As such, there is a trivial upper bound on the query complexity for classical algorithms: any graph problem can be solved classically with at most $O(|E|)$ queries, i.e., by accessing the whole input. This says nothing about the extra number of operations required. Besides queries to the input, we also comment on the extra number of operations required in classical algorithms and the extra number of simple gates required in the quantum algorithm. Nevertheless, we focus our comparison on the query complexities.

\section{Classical Sampling Algorithms}
\label{sec:classicallp}

\subsection{$A^2$ and $A^3$ Sampling Algorithms}

As discussed in Section \ref{sec:pathlp} some of the most basic but popular path-based link prediction methods are based on computing $A^2$ and $A^3$ to quantify the number of paths of length 2 and 3, respectively, between pairs of nodes. Here, we study classical algorithms to sample entries from $f(A)=A^2$ and $f(A)=A^3$. Algorithms \ref{A2samp_main} and \ref{A3samp_main} accomplish exactly that, as discussed in Appendix \ref{si:matrixpowers}. These are known classical routines, previously presented in Refs. \cite{seshadhri2013triadic} and \cite{jha2015path}. Our contribution here is simply their inclusion in our link prediction sampling framework and the study of their query complexity. These algorithms require an initial probability distribution to be processed and are then able to produce samples of links $(i,j)$ with probability
\begin{equation}
    \mathcal{P}[(i,j),\,A,\,n]=\frac{|(A^n)_{ij}|}{\|A^n\|_{1,1}}
\end{equation}
for $n=2$ and $3$. 

\begin{algorithm}[t]
\SetAlgoLined
\KwData{Graph $\mathcal{G}(V,\,E)$ in the GGM, integer $n_s$}
\KwResult{$n_s$ samples of pairs $(i,\,j)$}
\ForEach{$v\in V$}{compute $\tilde{p}_v=k_v^2$}
normalize $p_v=\tilde{p}_v/\sum_{v}\tilde{p}_v$\\
s = 0\\
\While{$s<n_s$}{
select $v\in V$ with probability $p_v$\\
randomly select $i\in\Gamma(v)$\\
randomly select $j\in\Gamma(v)$\\
\If{$i\neq j$ and $(i,\,j)\notin E$}{save $(i,\,j)$, $s=s+1$}}
\Return{\normalfont list of pairs $(i,\,j)$}
\caption{$A^2$ Sampling, adapted from \cite{seshadhri2013triadic}}
\label{A2samp_main}
\end{algorithm}

\begin{algorithm}[t]
\SetAlgoLined
\KwData{Graph $\mathcal{G}(V,\,E)$ in the GGM, integer $n_s$}
\KwResult{$n_s$ samples of pairs $(i,\,j)$}
\ForEach{$v\in V$}{
    \ForEach{$u\in\Gamma(v)$}{
        compute $\tilde{p}_{uv}=k_uk_v$
    }
}
normalize $p_{uv}=\tilde{p}_{uv}/\sum_{uv}\tilde{p}_{uv}$\\
s = 0\\
\While{$s<n_s$}{
select (u, v) with probability $p_{uv}$\\
randomly select $i\in\Gamma(u)$\\
randomly select $j\in\Gamma(v)$\\
\If{$i\neq j$ and $(i,\,j)\notin E$}{save $(i,\,j)$, $s=s+1$}}
\Return{\normalfont list of pairs $(i,\,j)$}
\caption{$A^3$ Sampling, adapted from \cite{jha2015path}}
\label{A3samp_main}
\end{algorithm}

\subsection{Complexity Analysis}

To compute the initial distributions $p_v$ and $p_{uv}$, Algorithm \ref{A2samp_main} queries the degree of each node $v\in V$ in lines 1-3, having query complexity $O(N)$, and \ref{A3samp_main} queries the degree and neighbours of each node $v\in V$ in lines 1-5, having query complexity $O(|E|)$. For each query one additional operation is used to compute the entries of $p_v$ and $p_{uv}$, and thus each algorithm requires an additional $O(N)$ and $O(|E|)$ operations, respectively.

Once the distributions $p_v$ and $p_{uv}$ are computed, Algorithms \ref{A2samp_main} and \ref{A3samp_main} can produce multiple samples of links $(i,\,j)$, as described in lines 6-13 and 8-15, respectively. First, algorithm \ref{A2samp_main} samples a node $v$ with probability $p_v$ in line 7, and Algorithm \ref{A3samp} samples a link $(u,\,v)$ with probability $p_{uv}$ in line 9. This requires a processing of $p_v$ and $p_{uv}$ into a cumulative array, requiring $O(N)$ and $O(|E|)$ operations, respectively, and then an additional $O(\log N)$ and $O(\log |E|)$ operations to bisect the array and draw each sample, as discussed in Appendix \ref{si:matrixpowers}. No additional queries to the input are required.

Finally, having sampled $v$ or $(u,\, v)$, a link $(i,\,j)$ is sampled by randomly selecting nodes from the neighbourhood of $v$ in lines 8-9 of Algorithm \ref{A2samp}, or from the neighbourhood of $(u,\,v)$ in lines 10-11 of Algorithm \ref{A3samp}, and then checking if $(i, j)$ is a useful sample in lines 10-12 and 12-14, respectively. We note also that this final step takes no additional queries to the input in Algorithm \ref{A3samp}, as the whole graph has already been queried during the processing of $p_{uv}$. However, for Algorithm \ref{A2samp}, the only information queried so far is the degree of each node. Each sample requires two extra neighbourhood queries to learn $i$ and $j$, and one extra vertex-pair query to learn if $(i,\,j)\in E$. This will lead to approximately $O(1/p_\text{G})$ queries per useful sample, or an extra $O(n_s/p_\text{G})$ queries in total, assuming small enough $n_s$. The higher the $n_s$ the higher the chance of drawing repeated samples, which require no extra queries, and eventually converging to the maximum number of queries $O(|E|)$.

In summary, drawing $n_s$ samples of useful links $(i,\,j)$, i.e., with $A_{ij}=0$ and $i\neq j$, following Algorithm \ref{A2samp} costs 
\begin{equation}
O\left(N+\frac{n_s}{p_\text{G}}\right)
\end{equation}
queries to the input, and takes an extra
\begin{equation}
O\left(N+\frac{n_s}{p_\text{G}}\log N\right)
\end{equation}
operations. Drawing $n_s$ samples of useful links $(i,j)$ following Algorithm \ref{A3samp} costs 
\begin{equation}
O(|E|)
\end{equation}
queries to the input and takes an extra
\begin{equation}
O\left(|E|+\frac{n_s}{p_\text{G}}\log|E|\right)
\end{equation}
operations.

The main takeaway here is that classical algorithms access the input a number of times that scales linearly with the input size $N$. We make this simplified statement as in complex networks the difference between $O(N)$ and $O(|E|)=O(Nk_\text{av})$ is often small due to the low average connectivity $k_\text{av}\ll N$ \cite{barabasi2016network}. While this cost is mostly due to the need to pre-compute $p_v$ and $p_{uv}$ before drawing samples, even if $p_v$ and $p_{uv}$ were given in the input model a similar query cost would be required to prepare the cumulative arrays for efficient bisection, described in Appendix \ref{si:matrixpowers}.

Our main objective now will be to show that a quantum algorithm can produce path-based link prediction samples using a quantum walk model with a number of input queries that is sub-linear in $N$.

\section{Quantum Link Prediction}
\label{sec:quantum}

\subsection{An Improved Algorithm for Link Prediction}

Recently, in Ref.\ \cite{moutinho2021quantum}, a link prediction method was proposed using continuous-time quantum walks to encode predictions based on both even-length and odd-length paths. In the original work the algorithm proposed to implement this method characterizes the predictions associated with each node $j$ separately, thus requiring $N$ repetitions to characterize predictions over the whole network, necessarily leading to an $O(N)$ factor in the complexity. 

\begin{figure}[t]
    \centering
    \includegraphics[width=\columnwidth]{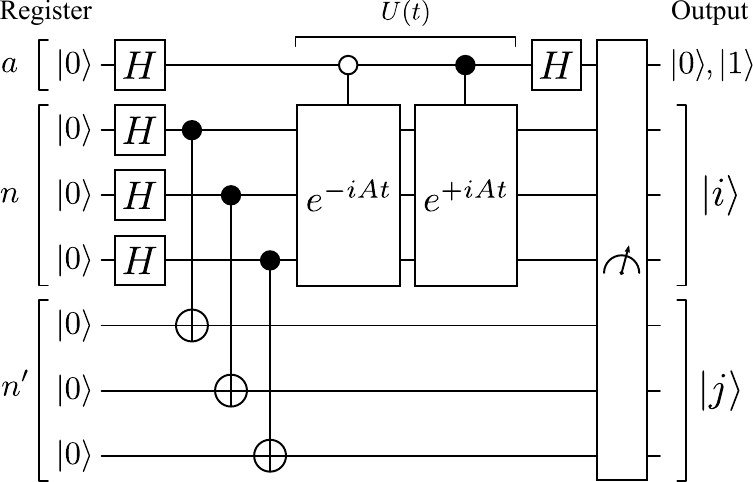}
    \caption{\textbf{QLP Circuit.} Example circuit to perform link prediction on a network with up to $N=8$ nodes. The total number of gates for a network with size $N$ is $\log_2 N+2$ Hadamard gates and $\log_2 N$ CNOT gates, plus the cost of implementing $U(t)$, described as $(e^{-iAt})_n$ conditional on $a=\ket{0}$ and $(e^{+iAt})_n$ conditional on $a=\ket{1}$.}
    \label{fig:qlpcircuit}
\end{figure}

Here we provide an improved quantum link prediction algorithm by designing it in such a way that links can be sampled globally from the network, without the need to fix an initial node. To do so, we consider a total of $2\log_2N + 1$ qubits: a register $n$ with $\log_2N$ qubits to represent each basis state $\ket{j}$ corresponding to a localized state at a node $j$ in the network, an extra register of qubits $n'$ with the same size as the register of node qubits $n$, and one ancilla register $a$ with a single qubit. We proceed now with the description of the circuit, exemplified in Fig.\ \ref{fig:qlpcircuit}. All qubits are initialized in the $\ket{0}$ state,
\begin{equation}
\ket{0}_a\ket{0}_n\ket{0}_{n'},
\end{equation}
after which Hadamard gates are applied to both the ancilla qubit and the register $n$ leading to,
\begin{equation}
\frac{1}{\sqrt{2}}(\ket{0}+\ket{1})_a\left(\frac{1}{\sqrt{N}}\sum_{w=1}^N\ket{w}\right)_n\ket{0}_{n'}.
\end{equation}
By applying a CNOT gate between each qubit in $n$ and the respective duplicate in $n'$, we effectively prepare the state
\begin{equation}
\frac{1}{\sqrt{2}}(\ket{0}+\ket{1})_a\frac{1}{\sqrt{N}}\sum_{w=1}^N\ket{w}_n\ket{w}_{n'}. \label{eq:initialstate}
\end{equation}
The quantum walk is now performed on register $n$ with an ancilla-controlled operator, while the register $n'$ remains unchanged. Consider then an operator $U(t)$ performing this quantum walk for some time $t$,
\begin{equation}
U(t)=\ketbra{0}{0}_a\left(e^{-iAt}\right)_nI_{n'}+\ketbra{1}{1}_a\left(e^{+iAt}\right)_nI_{n'},
\end{equation}
where $I_{n'}$ is the identity operator on register $n'$. Applying this operator to state \ref{eq:initialstate} leads to
\begin{equation}
\begin{split}
\ket{\psi(t)}=\frac{1}{\sqrt{2N}}&\left[\ket{0}_a\sum_{w=1}^N\left(e^{-iAt}\right)_n\ket{w}_n\ket{w}_{n'}\right.\\
&\left.+\ket{1}_a\sum_{w=1}^N\left(e^{+iAt}\right)_n\ket{w}_n\ket{w}_{n'}\right].
\end{split}
\end{equation}
With a final Hadamard gate on register $a$ the $\ket{0}_a$ and $\ket{1}_a$ subspaces interfere, leading to a sum of the exponential terms for $\ket{0}_a$ and a subtraction for $\ket{1}_a$, which we rewrite as the cosine and sine functions,
\begin{equation}
\begin{split}
\ket{\psi(t)}=&\ket{0}_a\left[\frac{1}{\sqrt{N}}\sum_{w=1}^N\cos(At)_n\ket{w}_n\ket{w}_{n'}\right]\\
+&i\ket{1}_a\left[\frac{1}{\sqrt{N}}\sum_{w=1}^N\sin(At)_n\ket{w}_n\ket{w}_{n'}\right].
\end{split}
\label{eq:qlpstate}
\end{equation}

From state \ref{eq:qlpstate} links can be sampled by measuring all qubits in the computational basis, as we now describe. The first step is to measure the ancilla qubit, yielding $\ket{0}_a$ or $\ket{1}_a$ with probabilities
\begin{align}
p^\text{even}(t)=\left\|\frac{1}{\sqrt{N}}\sum_{w=1}^N\cos(At)_n\ket{w}_n\ket{w}_{n'}\right\|^2,\\
p^\text{odd}(t)=\left\|\frac{1}{\sqrt{N}}\sum_{w=1}^N\sin(At)_n\ket{w}_n\ket{w}_{n'}\right\|^2,
\end{align}
respectively. By subsequently measuring the registers $n$ and $n'$ the state of these qubits will collapse to some $\ket{i}_n\ket{j}_{n'}$ basis state, corresponding to a sample of a link $(i,\,j)$. Samples corresponding to link predictions based on even or odd paths can be post-selected depending on register $a$ being $\ket{0}_a$ or $\ket{1}_a$, respectively. For both cases, the probability that some link $(i,\,j)$ is sampled can be computed by projecting the $\ket{0}_a$ or $\ket{1}_a$ component of Eq. \ref{eq:qlpstate} onto $\ket{i}_n\ket{j}_{n'}$, leading to
\begin{align}
p_{ij}^\text{even}(t)&=\frac{1}{N}|\mel{i}{\cos(At)}{j}|^2\label{eq:evenprob},\\
p_{ij}^\text{odd}(t)&=\frac{1}{N}|\mel{i}{\sin(At)}{j}|^2\label{eq:oddprob}.
\end{align}

The values of $p_{ij}^\text{even}(t)$ and $p_{ij}^\text{odd}(t)$ represent the even and odd path-based prediction scores that are coded into the quantum walk through the power series of the cosine and sine functions. These are the same prediction scores obtained in the original QLP method from \cite{moutinho2021quantum} with an extra $1/N$ normalizing factor.

From this point forward we will use these probabilities represented by the entries of the $\cos(At)$ and $\sin(At)$ matrices,
\begin{align}
    p_{ij}^\text{even}(t)&=\frac{1}{N}|\cos(At)_{ij}|^2,\label{eq:probs1}\\
    p_{ij}^\text{odd}(t)&=\frac{1}{N}|\sin(At)_{ij}|^2,\label{eq:probs2}\\
    p^\text{even}(t)&=\sum_{ij}\nolimits^Np_{ij}^\text{even}(t),\label{eq:probs3}\\
    p^\text{odd}(t)&=\sum_{ij}\nolimits^Np_{ij}^\text{odd}(t),\label{eq:probs4}\\
    p^\text{even}(t)&+p^\text{odd}(t)=1\label{eq:probs5}
\end{align}

\begin{algorithm}[t]
\SetAlgoLined
\KwData{Graph $\mathcal{G}(V,\,E)$, integer $n_s$}
\KwResult{$n_s$ samples of pairs $(i,\,j)$}
\While{$s<n_s$}{
apply circuit in Fig.\ \ref{fig:qlpcircuit}\\
\If{$a=\ket{0}$, $i\neq j$ and $(i,\,j)\notin E$}{save $(i,\,j)$ as even prediction, $s=s+1$\\}
\If{$a=\ket{1}$, $i\neq j$ and $(i,\,j)\notin E$}{save $(i,\,j)$ as odd prediction, $s=s+1$\\}
}
\caption{QLP Sampling}
\label{qlp_alg}
\end{algorithm}

In summary, the circuit in Figure \ref{fig:qlpcircuit} starts from an equal superposition of all nodes in the graph as in Eq. \ref{eq:initialstate}, evolves it according to a controlled quantum walk leading to Eq. \ref{eq:qlpstate}, and measures a basis state $\ket{0}_a\ket{i}_n\ket{j}_{n'}$ or $\ket{1}_a\ket{i}_n\ket{j}_{n'}$ corresponding to even or odd path-based sample of a link $(i,\,j)$, with probabilities given by the expressions $p_{ij}^\text{even}(t)$ and $p_{ij}^\text{odd}(t)$, respectively.

Simulations done in Ref.\ \cite{moutinho2021quantum} have already shown that a classical simulation of the quantum walk process, with the probability distributions being directly used as scores for link prediction, produces results with good prediction precision when compared with other state-of-the-art classical path-based methods over a wide range of real complex networks. In the following sections we consider the improved QLP Algorithm presented here, summarized in Algorithm \ref{qlp_alg}, and study how an actual quantum computing implementation would scale in terms of query access to the input to produce a fixed number of link prediction samples.

\subsection{Complexity Analysis}
\label{sec:qlpcomplexity}

Considering now Algorithm \ref{qlp_alg} using the circuit exemplified in Fig.\ \ref{fig:qlpcircuit}, we describe the resources required in terms of query access to the input. To produce a sample $(i,\,j)$ given a network with $N$ nodes, other than the application of $U(t)$, the circuit in Fig.\ \ref{fig:qlpcircuit} uses $\log_2N+2$ Hadamard gates and $\log_2 N$ CNOT gates. The main computational cost of the circuit is indeed the application of $U(t)$, which can be described as two applications of $e^{\pm iAt}$ conditioned on the ancilla qubit. Having an efficient implementation of $e^{-iAt}$ in a quantum computer for some hermitian matrix $A$ is the important problem of Quantum Simulation. It has been shown that for some classes of matrices it is possible to efficiently simulate $A\in\mathbb{R}^{N\times N}$ with $O(\text{polylog}(N))$ queries to the input \cite{lloyd1996universal, low2017optimal}. Until recently, the networks that were known to be efficiently simulatable were typically sparse, with all nodes having at most \text{polylog}$(N)$ connections. Real networks, however, have complex structural properties, including densely connected hubs with \text{poly}$(N)$ connections, community structures, small-world properties, and others. Finding a single efficiently simulatable model for complex networks in general may not be possible \cite{childs2009limitations}. However, motivated by complex network analysis, a recent study \cite{magano2022simulation} has shown that by adding polylog$(N)$ hub-nodes with $O(N)$ connections to an otherwise sparse network, the resulting adjacency matrix remains simulatable in $O(\text{polylog}(N))$. This toy-model for hub-sparse networks captures the important property of densely connected nodes in complex networks, and may inspire further research into the area of complex network simulation.

Let us start by describing a general query cost of implementing $e^{-iAt}$ as requiring $C$ queries to the input model in Section \ref{sec:input}. Then, to produce $n_s$ samples of useful links from the QLP algorithm, irrespectively of those being based on even or odd paths, we will need to run the circuit in Fig.\ \ref{fig:qlpcircuit} $O(n_s/p_\text{G})$ times to guarantee samples are useful, leading to a total of
\begin{equation}
    O\left(\frac{n_s}{p_\text{G}}C\right).
\end{equation}
queries to the input. Here, the $1/p_\text{G}$ factor conditioning the sample to be useful is multiplicative in the query complexity, as quantum algorithms to simulate $e^{-iAt}$ require the input to be queried in superposition. This information is ultimately lost when the final measurements produce each sample, meaning that for each desired sample the $C$ queries required to simulate $e^{-iAt}$ must be repeated.

To continue our discussion, it is useful to select a specific simulation algorithm so that we can provide a more concrete estimate for the resources of QLP in a general setting. We can consider, for example, the quantum simulation of $d$-sparse matrices, i.e., each row and column having at most $d$ elements, which can always be applied to any complex network at the cost of dealing with $d\sim\text{poly}(N)$. In Ref.\ \cite{low2017optimal} an optimal quantum algorithm to simulate $d$-sparse matrices was proposed that scales as
\begin{equation}
    C=O\left(td\|A\|_\text{max}+\text{polylog}(1/\epsilon)\right)
\end{equation}
in the number of queries, where $t$ is the time of the evolution, $\|A\|_\text{max}$ is the maximum entry of $A$ in absolute value and $\epsilon$ is the allowed error. In the original work it is considered that $d=O(\text{polylog}(N))$, and thus it is concluded that $d$-sparse matrices can be efficiently simulated. In our case, analysing QLP under the $d$-sparse model implies $d=k_\text{max}$, the maximum degree of the network, and $\|A\|_\text{max}=1$. We may then write the query complexity of QLP using the $d$-sparse model and disregarding the polylogarithmic factor on the error as
\begin{equation}
    \tilde{O}\left(\frac{n_s}{p_\text{G}}k_\text{max}t\right).
\end{equation}
As mentioned, one of the main structural properties of complex networks is the existence of large hubs. For complex networks described by a scale-free model the largest node degree in the network is estimated as
\begin{equation}
    k_\text{max}=O\left(N^\frac{1}{\gamma-1}\right)\label{eq:kmax}
\end{equation}
where $\gamma$ characterizes the power-law that describes the degree distribution, $k^{-\gamma}$. Typical values of $\gamma$ for real networks are in the $2<\gamma<4$ range \cite{barabasi2016network}. Given Eq. \ref{eq:kmax}, using the $d$-sparse model for the simulation of $A$ representing a scale-free complex network limits QLP to be at most polynomially faster than classical algorithms in the dependence on $N$. Nevertheless, this is sufficient for the analysis in the remainder of this work, and the possibility of an exponential speedup remains open given any future developments on the efficient simulation of complex networks.

\begin{table}[t]
    \centering
    \begin{tabular}{cc}
    \hline\hline
    Method  & Queries \\\hline
    $A^2\quad$ & $O\left(N+\frac{n_s}{p_\text{G}}\right)$ \\
    $A^3\quad$ & $O(|E|)$ \\\hline
    QLP (general)$\quad$& $\tilde{O}\left(\frac{n_s}{p_\text{G}}C\right)$\\
    QLP ($d$-sparse) $\quad$& $\tilde{O}\left(\frac{n_s}{p_\text{G}}k_\text{max}t\right)$\\
    \hline\hline
    \end{tabular}
    \caption{\textbf{Query complexity comparison.} For QLP we consider both a general algorithm using $C$ queries to implement $e^{-iAt}$ as well as the algorithm from Ref.\ \cite{low2017optimal} for $d$-sparse matrices that implements $e^{-iAt}$ with $\tilde{O}(k_\text{max}t)$ queries.}
    \label{tab:complexitycomparison}
\end{table}

We have just described the resources required to produce $n_s$ useful link samples from $A^2$ and $A^3$ using classical algorithms, and from QLP, a quantum algorithm encoding a series of even or odd powers of $A$, summarized in Table \ref{tab:complexitycomparison}. As mentioned, the query complexity of classical methods saturates at $O(|E|)$, after which no more queries are required as the whole graph has been read to memory. To proceed with our analysis we are going to focus on the an application of QLP using the $d$-sparse model from Ref.\ \cite{low2017optimal}. Here the query complexity is multiplicative in several parameters: the number of desired useful samples $n_s$, the number of samples per useful sample $1/p_\text{G}$, the time of the quantum walk $t$, and the maximum degree of the network $k_\text{max}$. We now wish to characterize the dependence on these parameters in order to comment on an achievable quantum speedup on the dependence on $N$, the size of the network.

\subsubsection{$1/p_\text{G}$ is a constant overhead}

\begin{figure}[ht!]
    \centering
    \includegraphics[width=0.9\columnwidth]{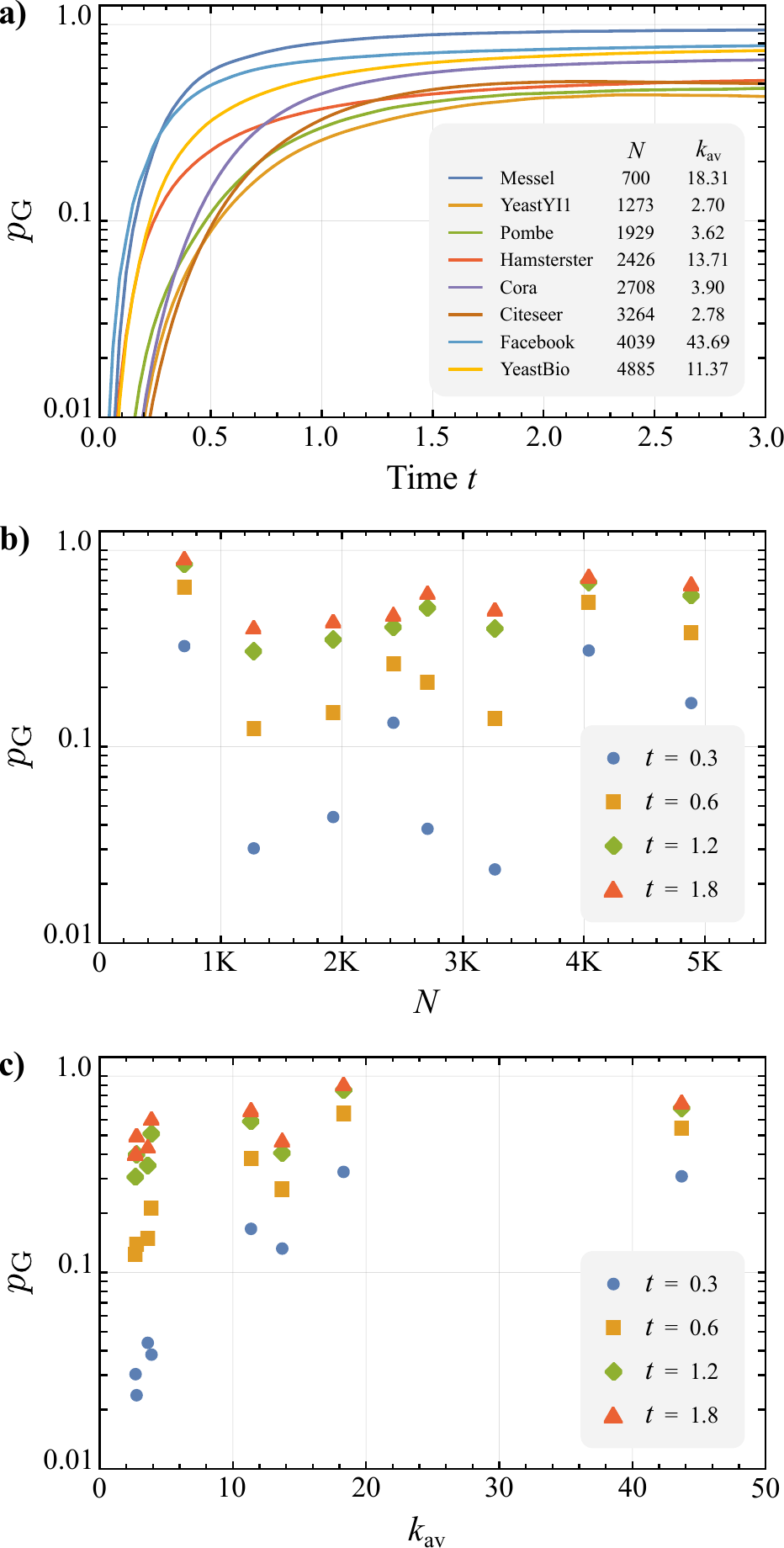}
    \caption{\textbf{Probability of sampling a useful link in QLP.} Simulating QLP on a range of datasets from real-world complex networks \cite{messel, biogrid, hamsterster, citeseercora, facebook} we computed the total probability of obtaining a useful sample, $p_\text{G}$. \textbf{a)} The probability saturates in the 0.1 to 1.0 range for increasing $t$, indicating it does not decrease with $N$. This is more explicit in the scatter plot \textbf{b)} of $p_\text{G}$ vs $N$ for different values of the time $t$ of the quantum walk. In \textbf{c)} we note there is a tendency for the useful link probability to increase with the average connectivity of the network, at least for low values of $k_\text{av}$. These results indicate the probability of sampling a useful link will typically be a constant overhead in the algorithm as $N$ increases.}
    \label{fig:realnetworks}
\end{figure}

To study the dependence of the query complexity of QLP on $1/ p_\text{G}$ we computed this probability over time for a range of real-world complex networks \cite{messel, biogrid, hamsterster, citeseercora, facebook} and synthetic networks. From Eqs. \ref{eq:pgood}, \ref{eq:probs1} and \ref{eq:probs2}, considering both even and odd-based samples, $p_\text{G}$ can be written as a time-dependent function on the time $t$ of the quantum walk,
\begin{equation}
    p_\text{G}(t) =\frac{1}{N}\sum_{i,j}^NG_{ij}\left(|\cos(At)_{ij}|^2+|\sin(At)_{ij}|^2\right),\label{eq:pgoodqlp}
\end{equation}
with $G$ the matrix of good prediction indices for a given complex network as defined in Eq. \ref{eq:matrixgood}. In Figure \ref{fig:realnetworks} we plot $p_\text{G}(t)$ for a range of real-world complex networks with different sizes and find that for all networks studied the value of $p_\text{G}(t)$ saturates in the $0.1$ to $1.0$ range as $t$ increases. Through the scatter plots shown in panels b) and c) we find that $p_\text{G}(t)$ does not decrease with $N$ and increases for small values of $k_\text{av}$ after which it remains approximately constant. Similar results were observed by repeating the same analysis in three different models of synthetic networks, as shown in Figs. \ref{fig:syntheticN} and \ref{fig:synthetick} of the Appendix. Here, we found that for all three models $p_\text{G}(t)$ remains exactly constant as $N$ increases, being only dependent on variations of the average degree, with a similar behaviour to that observed in real-world networks.

Overall, these results indicate that the number of required samples before observing a useful sample, $O(1/p_\text{G})$, can be considered as a constant overhead in the query complexity of QLP as $N$ increases. How large of an overhead will depend on the network, as well the value chosen for $t$, as shown in Fig.\ \ref{fig:realnetworks} a). However, as we will see in the next section, $t$ can typically be chosen such that $O(t/p_\text{G})$ is a small overhead while maintaining competitive precision in the method.

\subsubsection{$t$ is a constant overhead}

\begin{figure*}[t]
    \centering
    \includegraphics[width=1.0\textwidth]{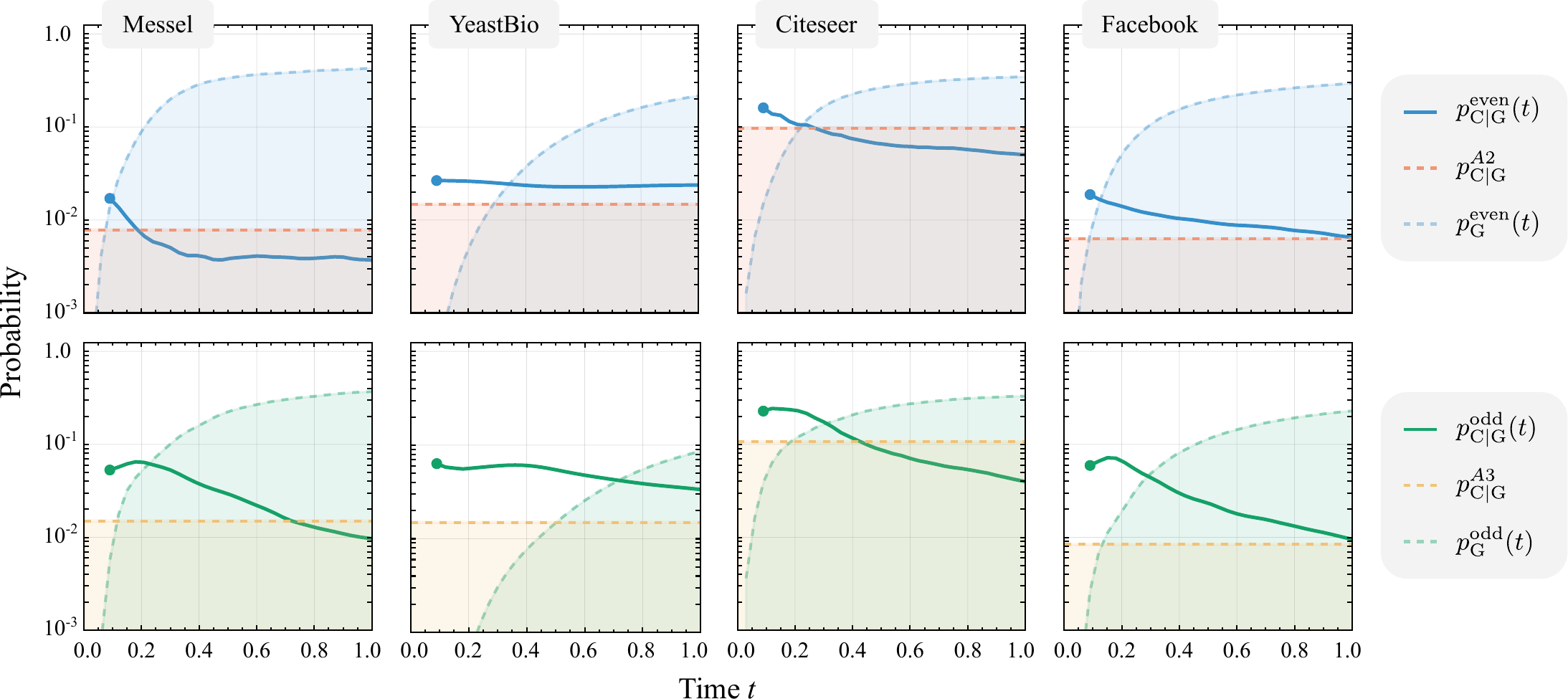}
    \caption{\textbf{Probability of sampling a correct link in QLP compared to $A^2$ and $A^3$.} We compare the precision of QLP with $A^2$ and $A^3$ by comparing the probability of sampling a correct prediction given that the sample was useful for four example networks. The results indicate that a value $t=O(1)$ can be chosen such that $p_{C|G}^\text{even}(t)\geq p_{C|G}^\text{A2}$ and $p_{C|G}^\text{odd}(t)\geq p_{C|G}^\text{A3}$ while maintaining a small useful sample overhead given by $1/p_\text{G}$. The plots for $p_{C|G}^\text{even}(t)$ and $p_{C|G}^\text{odd}(t)$ start at $t\approx0.1$ to avoid the region of small $p_\text{G}$ where the divisions in Eq. \ref{eq:pcorrectgood} are numerically unstable. Further results for more networks are provided in Fig.\ \ref{fig:precisionsi} of the Appendix.}
    \label{fig:precisionmain}
\end{figure*}

Next in our analysis is $t$, the time for which the quantum walk evolves over the network before each sample is obtained. This parameter influences both complexity and precision over three factors:

\textbf{Factor 1} - If the value chosen is too small the quantum walk does not spread significantly over the network and thus the probability of obtaining samples that are useful for link prediction is low, increasing the overhead of $O(1/p_\text{G})$ in the query complexity.

\textbf{Factor 2} - At the same time, higher values of $t$ imply that the quantum walk evolution must be simulated for longer, represented by the linear dependence with $t$ in the query complexity.

\textbf{Factor 3} - Finally, our goal is to do link prediction, and here $t$ acts as a hyper-parameter in the model which determines the weight of each power of $A$ in power series expansion of $\cos(At)$ and $\sin(At)$, which in turn influence the precision of the method.

With these three factors in mind, we wish to characterize how QLP behaves with changes in $t$ so that we can comment on its overall effect on the resources of the method. However, given the factors described, we can no longer focus our discussion solely on computational resources, but must also discuss precision. Given that we are considering classical algorithms based on $A^2$ and $A^3$, and a quantum algorithm that approximates matrix powers through $\cos(At)$ and $\sin(At)$, we need to guarantee that any claims we make on a resource advantage also admits a competitive precision. To analyse both resources and precision, we considered the following probabilities:
\begin{itemize}
    \item $p_\text{G}^\text{even}(t)$ \textbf{or} $p_\text{G}^\text{odd}(t)$ -- probability of obtaining a useful sample from either the $\cos(At)$ \textbf{or} $\sin(At)$ components of QLP.
    \item $p_\text{C|G}^\text{even}(t)$ \textbf{or} $p_\text{C|G}^\text{odd}(t)$ -- probability of obtaining a correct sample from QLP given that the sample was useful and obtained from $\cos(At)$ \textbf{or} $\sin(At)$, respectively.
    \item $p_\text{C|G}^{A2}$ \textbf{or} $p_\text{C|G}^{A3}$ -- probability of obtaining a correct sample given that the sample was useful and obtained from the respective classical algorithms for $A^2$ \textbf{or} $A^3$.
\end{itemize}

We start by commenting on $p_G^\text{even}(t)$ and $p_G^\text{odd}(t)$. These are the two contributions summing to the global probability of obtaining a useful sample from QLP $p_G(t)=p_G^\text{even}(t)+p_G^\text{odd}(t)$, as studied in the previous section,
\begin{align}
    p_\text{G}^\text{even}(t) = \frac{1}{N}\sum_{i,j}^NG_{ij}|\cos(At)_{ij}|^2,\label{eq:pgoodeven}\\
    p_\text{G}^\text{odd}(t) = \frac{1}{N}\sum_{i,j}^NG_{ij}|\sin(At)_{ij}|^2.\label{eq:pgoododd}
\end{align}

Consider now a matrix $A'$ encoding the solution to the link prediction problem, i.e., $A'_{ij}=1$ if $(i,j)$ is a correct prediction of a missing link in $A$. Then, the probability that a sample is a correct prediction based on the even or odd components of QLP is given by
\begin{align}
    p_\text{C}^\text{even}(t) = \frac{1}{N}\sum_{i,j}^NA'_{ij}|\cos(At)_{ij}|^2,\label{eq:pcorrecteven}\\
    p_\text{C}^\text{odd}(t) = \frac{1}{N}\sum_{i,j}^NA'_{ij}|\sin(At)_{ij}|^2.\label{eq:pcorrectodd}
\end{align}

Finally, the probabilities of obtaining a correct sample given that the sample was useful serve as a measure of precision of the method. For the quantum method, these are represented by $p_{C|G}^\text{even}(t)$ and $p_{C|G}^\text{odd}(t)$. 
\begin{equation}
    p_\text{C|G}^\text{even}(t) = \frac{p_\text{C}^\text{even}(t)}{p_\text{G}^\text{even}(t)},\qquad
    p_\text{C|G}^\text{odd}(t) = \frac{p_\text{C}^\text{odd}(t)}{p_\text{G}^\text{odd}(t)}. \label{eq:pcorrectgood}
\end{equation}
Similarly, $p_{C|G}^{A2}$ and $p_{C|G}^{A3}$ can be computed for each network with a respective adjacency matrix $A$ and a link prediction solution $A'$.

To characterize how QLP behaves with changes in $t$, we selected four example datasets where we compared the evolution of the listed probabilities over time for both even and odd-power results compared to the classical case of sampling from $A^2$ and $A^3$. For each dataset we performed a 10-fold cross validation procedure, where for each of the ten iterations $10\%$ of the links were randomly removed to build $A'$, and the prediction methods computed on the remaining $90\%$. In Fig.\ \ref{fig:precisionmain} we compare $p_\text{C|G}^\text{even}(t)$ with $p_{C|G}^{A2}$, superposed to $p_\text{G}^\text{even}(t)$, and $p_\text{C|G}^\text{odd}(t)$ with $p_{C|G}^{A3}$, superposed to $p_\text{G}^\text{odd}(t)$. Each result shown is an average over the ten iterations of the cross-validation procedure.

To get an intuitive reading of Fig.\ \ref{fig:precisionmain}, we start by noting that in all cases it is possible to pick a $t \lesssim 1$ such that $p_\text{G}^\text{even}(t)$ and $p_\text{G}^\text{odd}(t)$ are both greater than $0.1$, i.e., the sampling overhead to obtain a useful sample is small. Looking now at the precision comparison given by the $p_\text{C|G}$ curves, we note that for these values of $t$, both useful samples of the even and odd component of QLP tend to have a higher chance of being correct than those obtained from the classical algorithms for $A^2$ and $A^3$. This indicates that a value of $t=O(1)$ can typically be chosen such that QLP has competitive performance over classical sampling algorithms for $A^2$ and $A^3$ while maintaining a small sampling overhead given by $O(t/p_\text{G})$.

\section{Discussion and conclusions}
\label{sec:discussion}
To summarize our work, we have discussed sampling algorithms for path-based link prediction accessing the input network through the model described in Section \ref{sec:input}, and outputting samples of links $(i, j)$ following the distribution of scores given by a function of $A$. In the classical case, we have considered known algorithms to sample from $A^2$ and $A^3$ and concluded that they access the input network with a total number of queries that is linear in the number of nodes $N$,
\begin{equation}
    A^2\sim O\left(N+\frac{n_s}{p_G}\right),\quad A^3\sim O(Nk_\text{av})
\end{equation}

In the quantum case we have presented an improved version of the QLP algorithm from Ref.\ \cite{moutinho2021quantum} which can sample links globally from the network. The samples $(i, j)$ are drawn from a score distribution following $\cos(At)$ or $\sin(At)$, which through their power-series represent even and odd powers of $A$, respectively, weighted by the time $t$ of the quantum walk. Considering the $d$-sparse model for the quantum simulation of $e^{-iAt}$ we estimated that QLP has a query complexity of 
\begin{equation}
\tilde{O}\left(\frac{n_s}{p_\text{G}}k_\text{max}t\right),
\end{equation}
where $n_s $ is the total number of useful samples drawn, $p_\text{G}$ is the probability of obtaining a useful sample, $k_\text{max}$ is the maximum degree of the network, and $t$ is the time of the quantum walk. Through numerical simulations of QLP we concluded that it is possible to draw samples with competitive precision to those obtained from $A^2$ and $A^3$ with $t/p_\text{G}\sim O(1)$. As such, considering a direct comparison to the classical sampling algorithms following the score distributions from $A^2$ and $A^3$, our final estimate for the complexity of QLP is
\begin{equation}
    \tilde{O}(n_sk_\text{max})\label{eq:finalqlpcomplexity}
\end{equation}
in the number of queries to the input. If we consider networks following a scale-free model with a power-law degree distribution given by $k^{-\gamma}$, then 
\begin{equation}
    k_\text{max}=O\left(N^\frac{1}{\gamma-1}\right).
\end{equation}
Typical complex networks have a power law in the $2<\gamma\leq4$ range, implying that the resources described in Eq. \ref{eq:finalqlpcomplexity} are sub-linear in $N$ and constitute a polynomial speedup over the classical algorithms for $A^2$ and $A^3$, as long as
\begin{equation}
    n_s<N^\frac{\gamma-2}{\gamma-1}.
\end{equation}

We emphasize that the results described here are based on the quantum simulation algorithm from Ref.\ \cite{low2017optimal} for $d$-sparse matrices, which remains valid for any complex network at the cost of having $d\sim\text{poly}(N)$. Nevertheless, QLP is independent of the method used to simulate the quantum walk, and the possibility of an exponential speedup remains open given the discussion in Section \ref{sec:qlpcomplexity}. Depending on the structural properties found within different types of complex networks, there may be more efficient quantum simulation algorithms. For networks following the hub-sparse model described in \cite{magano2022simulation}, for example, the resources of QLP would scale as polylog$(N)$, and this would be constitute an exponential speedup over a classical implementation of $A^2$ or $A^3$ sampling in these same networks.

In regards to future directions, there are few options to consider. As mentioned in the introduction, several different methods exist for link prediction, and here we focused on path-based methods which have been shown to perform reasonably well in a wide range of network types \cite{zhou2021progresses, zhou2021experimental, muscoloni2022adaptive, muscoloni2021short}. For path-based link prediction, although we used a quantum walk formulation, a quantum algorithm using a direct implementation of adjacency matrix powers could also be possible, for example, using the formalism of Quantum Singular Value Transformations \cite{gilyen2019quantum}. Nevertheless, quantum algorithms for other link prediction approaches may also be developed with a potential for quantum advantage. One example would be to use quantum machine-learning techniques to adapt classical methods that learn an optimized prediction function \cite{ghasemian2020stacking}. In that case, an efficient quantum implementation of each predictor would be required.

\begin{acknowledgments}
The authors thank Yasser Omar for valuable discussions, and thank Funda\c{c}\~{a}o para a Ci\^{e}ncia e a Tecnologia (FCT, Portugal) for the support through project UIDB/EEA/50008/2020. Furthermore, JPM and DM acknowledge the support of FCT through scholarships 2019.144151.BD and 2020.04677.BD, respectively, and thank Yasser Omar for his supervision. BC acknowledges the support of FCT through project CEECINST/00117/2018/CP1495/CT0001.
\end{acknowledgments}

\bibliography{draft}
\bibliographystyle{unsrt}


\clearpage

\appendix
\section{Classical matrix power algorithms}
\label{si:matrixpowers}
\subsection{Deterministic Algorithms}
As discussed in the main text, path-based link prediction often requires computing powers of the adjacency matrix in order to find high-valued entries. In this section we discuss classical algorithms for this task, both deterministic and randomized. Let us start with the simple cases of computing $A^2$ or $A^3$. If $A$ is sparse enough, it may be advantageous to compute these matrices by explicitly counting all paths of length 2 and 3 in the graph, as written in Algorithms \ref{A2det} and \ref{A3det}.

\begin{algorithm}[t]
\SetAlgoLined
\KwData{Graph $\mathcal{G}(V,\,E)$}
\KwResult{$A^2$}
$(A^2)_{ij}=0$ \textbf{forall} $i,j=1,...,N$\\
\ForAll{$v\in\mathcal{V}$}{
 \ForAll{$i\in\Gamma(v)$}{
  \ForAll{$j\in\Gamma(v)$}{
	$(A^2)_{ij}=(A^2)_{ij}+1$
   }
  }
 }
 \caption{$A^2$ Counting\label{A2det}}
\end{algorithm}
\begin{algorithm}[t]
\SetAlgoLined
\KwData{Graph $\mathcal{G}(V,\,E)$}
\KwResult{$A^3$}
$(A^3)_{ij}=0$ \textbf{forall} $i,j=1,...,N$\\
\ForAll{$v\in\mathcal{V}$}{
 \ForAll{$u\in\Gamma(v)$}{
  \ForAll{$i\in\Gamma(v)$}{
   \ForAll{$j\in\Gamma(u)$}{
    $(A^3)_{ij}=(A^3)_{ij}+1$
   }
  }
 }
}
\caption{$A^3$ Counting\label{A3det}}
\end{algorithm}

For the sake of simplicity, we did not exploit the fact that $A$ is symmetric. Both of these algorithms must access the full graph, and thus they have query complexity $O(Nk_\text{av})=O(|E|)$, corresponding to the number of neighbour queries required to write the graph to memory. In addition to the number of queries, we can also characterize the time complexity given the number of extra operations required, corresponding to the nested loops that run through all paths of the given length in the graph. It was shown in Ref.\ \cite{fiol2009number} that
\begin{equation}
\sum_{i,j}^N(A^n)_{ij}\leq N\langle k^n\rangle \label{eq:apowercondition}
\end{equation} 
where the equality holds for $n=2$, and $\langle k^n\rangle$ is the $n$-th moment of the degree distribution in the graph,
\begin{equation}
\langle k^n\rangle = \frac{1}{N}\sum_{i=1}^Nk_i^n.
\end{equation}
From Eq. \ref{eq:apowercondition} it follows that the time complexity of Algorithm \ref{A2det} is $O(N\langle k^2\rangle)$, and $O(N\langle k^3\rangle)$ for Algorithm \ref{A3det}.

\subsection{Sampling Algorithms}

Instead of deterministically computing $A^n$ to find high-valued entries, however, we may consider algorithms that sample an entry $(i,j)$ of $A^n$ with probability
\begin{equation}
\mathcal{P}[(i,j),\,A,\,n]=\frac{|(A^n)_{ij}|}{\|A^n\|_{1,1}}
\end{equation}
where $\|.\|_{1,1}$ denotes the $L_{1,1}$ matrix norm. For $A^2$ and $A^3$ we write Algorithms \ref{A2samp} and \ref{A3samp}, inspired by Refs. \cite{seshadhri2013triadic} and \cite{jha2015path}, respectively.

Let us start by analysing Algorithm \ref{A2samp} to sample entries of $A^2$. The first step computes the distribution 
\begin{equation}
p_v=k_v^2/\sum_{v=1}^Nk_v^2
\end{equation}
over all nodes $v\in V$, so that we may then sample a node $v\in V$ from this distribution. The quantity $k_v^2$ counts the number of paths of length 2 going through node $v$, disregarding direction and repetition of starting and ending nodes. Then, a pair $(i, j)$ is chosen where both $i$ and $j$ are independent random samples from the neighbourhood of $v$. Thus, the probability to pick a certain $(i,j)$ conditional on $v$ is
\begin{equation}
\mathcal{P}[(i,j)|v]=\frac{1}{k_v}\frac{1}{k_v}
\end{equation}
Finally, the probability of sampling a pair $(i,j)$ can be computed by summing over all nodes $v$ in the graph,
\begin{align}
\mathcal{P}[(i,j)]&=\sum_{v=1}^N\mathcal{P}[v]\mathcal{P}[(i,j)|v]\\
&=\sum_{v\in\Gamma(i)\cap\Gamma(j)}\frac{k_v^2}{\sum_{v'}^Nk_{v'}^2}\frac{1}{k_v^2}\\
&=\frac{|\Gamma(i)\cap\Gamma(j)|}{\sum_{v'}^Nk_{v'}^2}\\
&=\frac{|(A^2)_{ij}|}{\|A^2\|_{1,1}},
\end{align}
where we defined the set of common neighbours between $i$ and $j$ as $\Gamma(i)\cap\Gamma(j)$, and used the following relations: $\mathcal{P}[(i,j)|v]=0$ for all $v\notin\Gamma(i)\cap\Gamma(j)$, $(A^2)_{ij}=|\Gamma(i)\cap\Gamma(j)|$, and $\sum_v^Nk_v^2=\sum_{i,j}^N(A^2)_{ij}=\|A^2\|_{1,1}$.

\begin{algorithm}[t]
\SetAlgoLined
\KwData{Graph $\mathcal{G}(V,\,E)$}
\KwResult{Pair (i, j)}
\lForEach{$v\in V$}{compute $\tilde{p}_v=k_v^2$}
normalize $p_v=\tilde{p}_v/\sum_{v}\tilde{p}_v$\\
select $v\in V$ with probability $p_v$\\
randomly select $i\in\Gamma(v)$\\
randomly select $j\in\Gamma(v)$\\
\caption{$A^2$ Sampling \label{A2samp} \cite{seshadhri2013triadic}}
\end{algorithm}
\begin{algorithm}[t]
\SetAlgoLined
\KwData{Graph $\mathcal{G}(V,\,E)$}
\KwResult{Pair (i, j)}
\ForEach{$v\in V$}{
    \lForEach{$u\in\Gamma(v)$}{
        compute $\tilde{p}_{uv}=k_uk_v$
    }
}
normalize $p_{uv}=\tilde{p}_{uv}/\sum_{uv}\tilde{p}_{uv}$\\
select (u, v) with probability $p_{uv}$\\
randomly select $i\in\Gamma(u)$\\
randomly select $j\in\Gamma(v)$
\caption{$A^3$ Sampling \label{A3samp} \cite{jha2015path}}
\end{algorithm}

Algorithm \ref{A3samp} follows a similar reasoning to sample entries from $A^3$. The first step computes the distribution
\begin{equation}
p_{uv}=k_uk_v/\sum_{v=1}^N\sum_{u\in\Gamma(v)}k_uk_v
\end{equation}
so that a link $(u,v)$ may be sampled according to this distribution. The quantity $k_uk_v$ counts the number of paths of length 3 going through link $(u,v)$, starting in a neighbour of $u$ and ending in a neighbour of $v$. Then, a pair $(i,j)$ is chosen where $i$ is a random sample from the neibhourhood of $u$ and $j$ is a random sample from the neighbourhood of $v$. The probability to pick a certain $(i, j)$ conditional on $(u, v)$ is
\begin{equation}
\mathcal{P}[(i,j)|(u,v)]=\frac{1}{k_uk_v}.
\end{equation}
We now define the set
\begin{equation}
L3(i,j)=\{(u,v):(i,u,v,j)\text{ is a path in }\mathcal{G}\},
\end{equation}
which implies that $(A^3)_{ij}=|L3(i,j)|$. Thus, following Algorithm \ref{A3samp}, the probability to sample a certain pair $(i,j)$ is
\begin{align}
\mathcal{P}[(i,j)]&=\sum_{(u,v)\in L3(i,j)}\mathcal{P}[(u,v)]\mathcal{P}[(i,j)|(u,v)]\\
&=\sum_{(u,v)\in L3(i,j)}\frac{1}{k_uk_v}\frac{k_uk_v}{\sum_{v'=1}^N\sum_{u'\in\Gamma(v')}k_{u'}k_{v'}}\\
&=\frac{|(A^3)_{ij}|}{\|A^3\|_{1,1}}.
\end{align}

To compute the initial distributions $p_v$ and $p_{uv}$, both Algorithms \ref{A2samp} and \ref{A3samp} must perform one degree query for each $v\in V$, having query complexity $O(N)$. Algorithm \ref{A3samp} must also access each neighbour of $v$ individually, raising the query complexity to $O(|E|)$. For each query one additional operation is used to compute the entries of $p_v$ and $p_{uv}$, and thus the time complexity for this step is the same as the query complexity.

\begin{table*}[t]
\centering
\begin{tabular}{cccc}
\hline\hline
 Complexity & & Deterministic & Randomized \\\hline
 \multirow{2}{*}{Queries} & $A^2$ &$O(|E|)$ & $O(N+n_s)$ \\
 &$A^3$ & $O(|E|)$ & $O(|E|)$ \\\hline
 \multirow{2}{*}{Operations}&$A^2$ & $O(N\langle k^2\rangle)$ & $O(N+n_s\log N)$ \\
 &$A^3$ & $O(N\langle k^3\rangle)$ & $O(|E|+n_s\log|E|)$\\\hline\hline
\end{tabular}
\caption{Complexity comparisons between deterministic and randomized algorithms for $A^2$ and $A^3$. Deterministic algorithms output the full matrix, while randomized output $n_s$ samples of links $(i, j)$ with probability proportional to $|(A^n)_{ij}|$.}
\label{tab:complexitycomparison2}
\end{table*}

Once the distributions $p_v$ and $p_{uv}$ are computed, Algorithms \ref{A2samp} and \ref{A3samp} can produce multiple samples of links $(i,j)$. First, algorithm \ref{A2samp} samples a node $v$ with probability $p_v$, and Algorithm \ref{A3samp} samples a link $(u,v)$ with probability $p_{uv}$. This can be described as the general problem of sampling an entry $i$ of a vector $p\in\mathbb{R}^{N'}$ with probability proportional to $|p_i|$. To solve it, we first build a vector $\hat{p}\in\mathbb{R}^{N'}$, corresponding to the cumulative sum of entries in $p$,
\begin{equation}
    \hat{p}_i=\sum_{j=0}^ip_j
\end{equation}
and then randomly pick a value $0\leq x \leq\hat{p}_{N'}$. Finally, we bisect $\hat{p}$ to find the smallest entry $i$ such that $x<\hat{p}_i$. The entry $i$ corresponds to our sample of $p$.

Building $\hat{p}$ requires $O(N')$ sums and bisecting $\hat{p}$ has an additive $O(\log N')$ cost. As such, processing $p_v$ and $p_{uv}$ in Algorithms \ref{A2samp} and \ref{A3samp} maintain their respective time complexities of $O(N)$ and $O(|E|)$. Then, producing each sample of $v$ and $(u,v)$ requires $O(\log N)$ and $O(\log |E|)$ operations, respectively. No additional queries to the input are required.

Finally, having sampled $v$ in Algorithm \ref{A2samp} and $(u, v)$ in Algorithm \ref{A3samp}, a link $(i,j)$ is sampled by randomly selecting nodes from the neighbourhood of $v$ or $(u, v)$, which we consider to be $O(1)$ operations. We note also that this final step takes no additional queries to the input in Algorithm \ref{A3samp}, as the whole graph has already been queried during the processing of $p_{uv}$. However, for Algorithm \ref{A2samp}, each sample of a link $(i,j)$ requires two queries to the neighbourhood of the selected $v$. Repeated samples may end up reading the whole graph, at which point no new queries are required.

In summary, drawing $n_s$ samples of links $(i,j)$ following Algorithm \ref{A2samp} costs $O(N+n_s)$ queries to the input, up to a maximum of $O(Nk_\text{av})=O(|E|)$, and takes $O(N+n_s\log N)$ operations. Drawing $n_s$ samples of links $(i,j)$ following Algorithm \ref{A3samp} costs $O(|E|)$ queries to the input and takes $O(|E|+n_s\log|E|)$ operations. We present in Table \ref{tab:complexitycomparison2} a comparison of both deterministic and randomized algorithms in terms of their query and time complexity.

\begin{figure*}[t]
    \centering
    \includegraphics[width=0.8\textwidth]{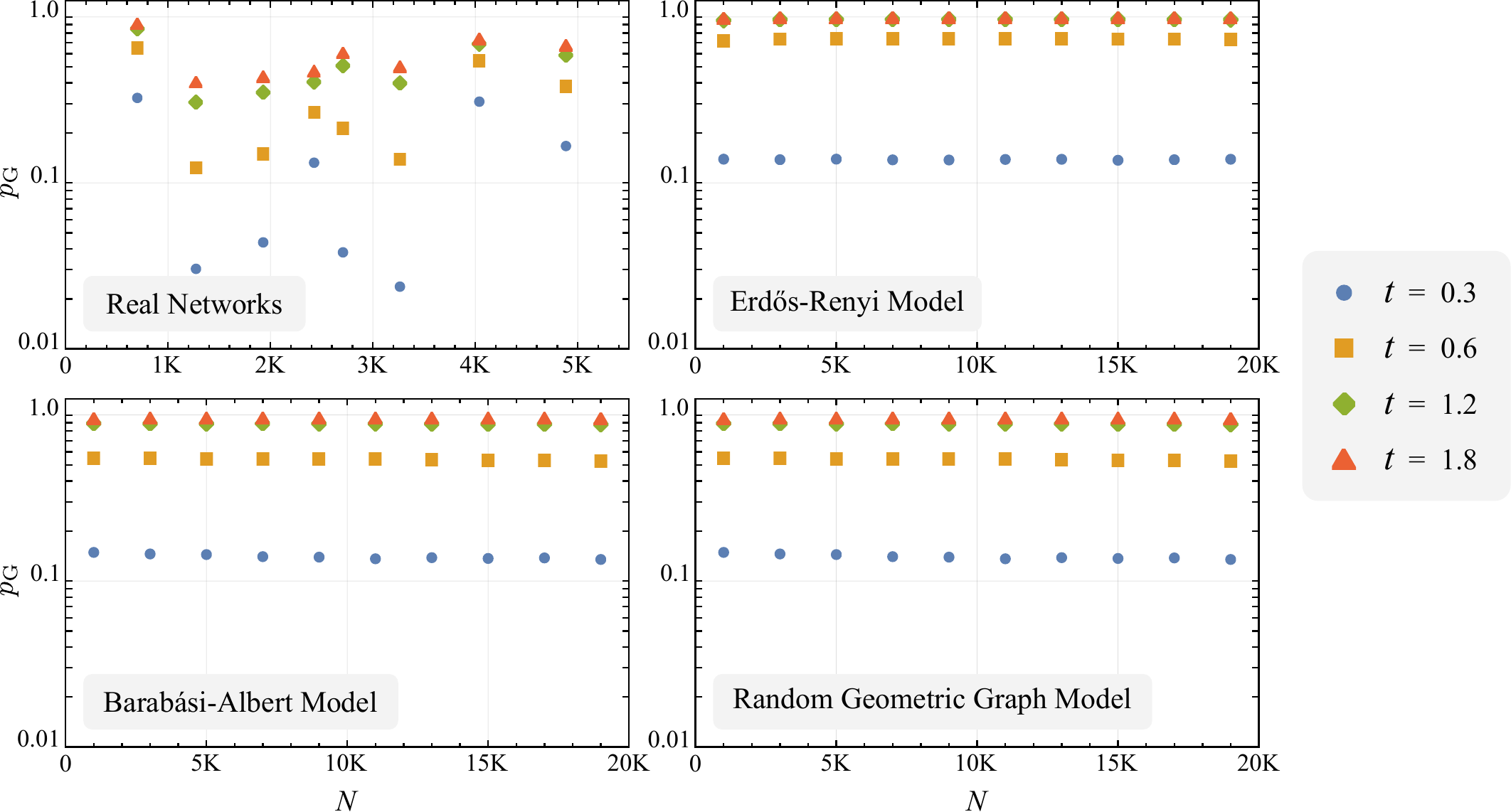}
    \caption{\textbf{Probability of sampling a useful link in QLP vs $N$.} We compare the variation in the probability of sampling a useful link with QLP with the size of the network $N$ in a range of real and synthetic networks. As $N$ increases, $p_\text{G}$ does not show a tendency to decrease in the real networks tested. To compare, we tested three different synthetic network models: the Erd\"{o}s-Renyi Model \cite{erdHos1959random}, the Barabási-Albert Model \cite{albert2002statistical}, and the Random Geometric Graph Model \cite{dall2002random}. In all cases we fixed $k_\text{av}=10$ and simulated QLP for different valuse of $t$, observing that $p_\text{G}$ remains exactly constant as $N$ increases.}
    \label{fig:syntheticN}
\end{figure*}

\begin{figure*}[t]
    \centering
    \includegraphics[width=0.8\textwidth]{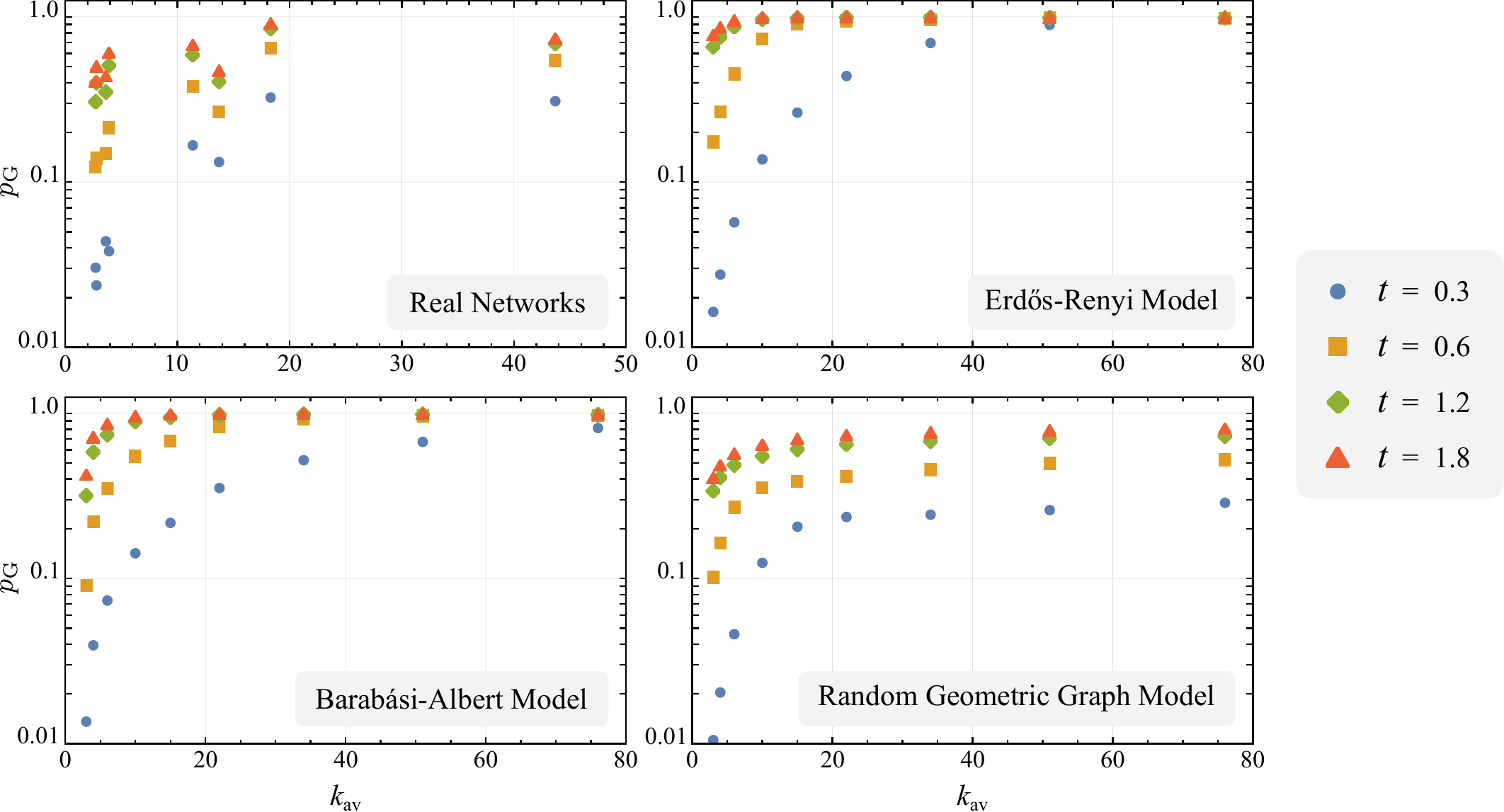}
    \caption{\textbf{Probability of sampling a useful link in QLP vs $k_\text{av}$.} Repeating the procedure of Fig.\ \ref{fig:syntheticN}, we fixed $N=5000$ and observed a similar behaviour in the synthethic and real networks: $p_\text{G}$ initially grows for increasing $k_\text{av}$, and then remains approximately constant.}
    \label{fig:synthetick}
\end{figure*}

\begin{figure*}[t]
    \centering
    \includegraphics[width=1.0\textwidth]{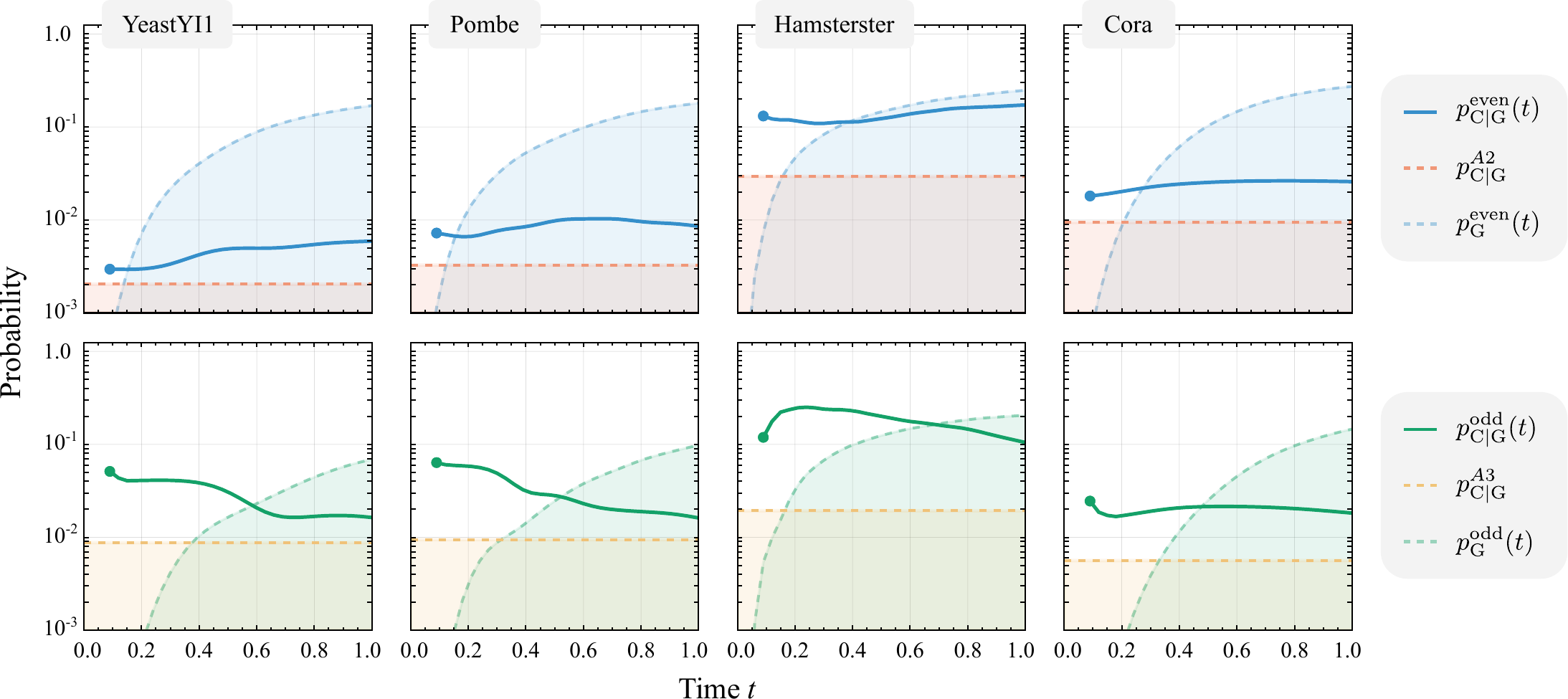}
    \caption{\textbf{Extra results for probability of sampling a correct link in QLP compared to $A^2$ and $A^3$.} We compare the precision of QLP with $A^2$ and $A^3$ by comparing the probability of sampling a correct prediction given that the sample was useful for four extra networks. The results indicate that a value $t=O(1)$ can be chosen such that $p_{C|G}^\text{even}(t)\geq p_{C|G}^\text{A2}$ and $p_{C|G}^\text{odd}(t)\geq p_{C|G}^\text{A3}$ while maintaining a small useful sample overhead given by $1/p_\text{G}$. The plots for $p_{C|G}^\text{even}(t)$ and $p_{C|G}^\text{odd}(t)$ start at $t\approx0.1$ to avoid the region of small $p_\text{G}$ where the divisions in Eq. \ref{eq:pcorrectgood} are numerically unstable.}
    \label{fig:precisionsi}
\end{figure*}

\end{document}